# ABOUT LINEAR THRESHOLD OF INSTABILITY OF FARADAY WAVES IN FLUID COVERED BY THIN FILM

E. Postnikov

The solution was found from Navier-Stokes equation and boundary conditions with interfacial tension as function of the film substance concentration. Here the method of Chen&Vinals is used. It was found that adding to coefficient leads to growth of critical value of the driving acceleration.

The Faraday waves are the clear example of self-organization in dynamical systems. See an overview of theoretical and experimental results of studying of this problem, for example, in [1-3].

There are the theoretical model for description of Faraday waves in case of the semi-infinite viscous fluid layer [3] and models for bounded depth [4], non-Newtonian liquid [5], two-frequency forced fluids, for example [6, 7]. The surface of liquid is clear in all these problems.

But there is a problem concerning fluid covered by thin liquid film in Faraday experiment. In known investigations one works with running waves, as long gravitational [8], as short capillary [9]. That makes hard experimental difficulties. The measurement of the threshold of Faraday instability is simpler but also it talks about mechanical and thermodynamical properties of the film.

Let one consider a semi-infinite fluid layer, unbounded in the x-y direction, extending to $z = -\infty$, with planar free surface at $z = 0$ whet at rest. The fluid is uncompressed and Newtonian, it has density $\rho$ and kinematic viscosity $\nu$. The fluid is covered by thin (near monomolecular) film with surface concentration $\gamma$.

The governing equations for fluid velocity **u** are

$$\frac{\partial \mathbf{u}}{\partial t} + (\mathbf{u} \cdot \nabla)\mathbf{u} = -\frac{1}{\rho}\nabla p + \nu \nabla^2 \mathbf{u} + \mathbf{g}(t) \qquad (1)$$

$$\operatorname{div} \mathbf{u} = 0,$$

where $\mathbf{g}(t) = (0,0,-g - f\cos\omega t)$.

The equation of continuity for liquid film is

$$\frac{\partial \gamma}{\partial t} + \operatorname{div}(\gamma \mathbf{u}) = 0. \qquad (2)$$

Besides there are boundary conditions

$$\mathbf{u} = 0 \text{ at } z = -\infty,$$

and at the free surface $z = \zeta$ [8,10]:

$$\mathbf{n}\cdot\mathbf{T}\cdot\mathbf{n} = 2H\sigma, \quad \mathbf{t}_m\cdot\mathbf{T}\cdot\mathbf{n} = -\frac{\partial \sigma}{\partial \gamma}\nabla\gamma\cdot\mathbf{t}_m, \ m=1,2 \ (x,y), \qquad (3)$$

where **T** is the stress tensor, H is the curvature of the surface, $\sigma$ is the interfacial tension, **t** and **n** are tangential and normal vectors.

The problem is to find linear threshold of instability. One can use for solution method from [3].

The vertical surface displacement and vertical velocity are

$$\zeta = \cos(kx) \sum_{j=1,3,5} e^{j\frac{i\omega t}{2}} A_j$$

$$w = u_z\big|_{z=\zeta} = \cos(kx) \sum_{j=1,3,5} e^{j\frac{i\omega t}{2}} w^j(z) A_j. \qquad (4)$$

In linear case $\mathbf{n} = (0,0,1)$, $\mathbf{t} = (1,1,0)$ and the concentration is $\gamma = \gamma_0 + \gamma'$, $\gamma_0 \gg \gamma'$, where $\gamma_0$ is constant concentration of the substance of film at surface at rest.

In this case one has from (2)

$$\frac{\partial \gamma'}{\partial t} - \gamma_0 \frac{\partial w}{\partial z} = 0. \qquad (5)$$

Boundary conditions are

$$\left[2\nu\nabla_H^2 - (\partial_t - \nu\nabla^2)\right]\partial_z w + \left(g - \frac{\sigma}{\rho}\nabla_H^2 + f\cos\omega t\right)\nabla_H^2 \zeta = 0, \qquad (6)$$

for normal stress and

$$\left(\nabla_H^2 - \partial_{zz}\right)\partial_t w = -\frac{\alpha\sigma}{\rho\nu}\nabla_H^2 \partial_z w, \qquad (7)$$

for tangential (using (5)).

Kinematic boundary condition is

$$\partial_t \zeta - w = 0. \qquad (8)$$

There is $\nabla_H^2 = \partial_{xx} + \partial_{yy}$ in (6)-(8). The coefficient $\alpha = (\gamma_0/\sigma)(\partial\sigma/\partial\gamma)$ has value from 0 to 1 and describes elastic properties of the film.

Substituting (4) into (7) and (8) and using the null conditions at $z = -\infty$ one finds

$$w^j(z) = \left(\frac{1}{2}ji\omega + 2\nu k^2 \beta\right)e^{kz} - 2\nu k^2 \beta e^{q_j z}, \qquad (9)$$

where $q_j^2 = k^2 + j\frac{i\omega}{2\nu}$, $\beta = \dfrac{1 - i\alpha\Sigma\dfrac{\omega}{4k^2\nu j}}{1 - \alpha\Sigma\dfrac{(1-q/k)}{j^2}}$, $\Sigma = \dfrac{4\sigma k^3}{\rho\omega^2}$.

Using (4), (6), (9) one finds set of equations like set in [3]:

$$H_1 A_1 - fA_1^* - fA_3 = 0,$$
$$H_3 A_3 - fA_1 - fA_5 = 0$$
$$H_5 A_5 - fA_3^* - fA_7 = 0$$
$$\ldots\ldots\ldots\ldots\ldots\ldots\ldots,$$

but it has other coefficients:

$$H_j = \frac{2}{k}\left\{v^2\left[4q_jk^4 - k(q_j^2 + k^2)\right] + 2v^2k^3(1-\beta)(k-q_j)^2 - gk^2 - \frac{\sigma}{\rho}k^2\right\}. \quad (10)$$

For a given wave number k, its threshold of instability $f_k$ is given like [3] by

$$f_k = \left|H_1 - \cfrac{f_k^2}{H_3 - \cfrac{f_k^2}{H_5 - ...}}\right|. \quad (11)$$

The critical wave number for instability corresponds to the lowest value of $f_k$. The existence of the film leads to the growth of critical value of the acceleration.

If $\alpha = 0$ (no film) $\beta = 1$ and (10) leads to known expression [3]. After truncation (11) at the first term one has one mode solution for running wave [8, 9].

REFERECES

1. Kudrolli A., Gollub J.R. Pattern and spatiotemporal chaos in parametrically forced surface waves: a systematic survey at large aspect ratio // Physica D.– V. 97.– 1996.– P. 133–154.
2. Gollub J.R. Pattern formation in nonequilibrium physics // Reviews of Modern Physics.– V. 71.– № 2.– 1999.– P. 386–403.
3. Chen P., Vinals J. Amplitude equation and pattern selection in Faraday waves // Physical Review E.– V. 60.– 1999.– P. 559–570
4. Binks D., Westra M.–T., van de Water W. Effect of dept on the formation of Faraday waves // Physical Review Letters.– V. 79.– 1997.– P. 5010–5013.
5. Wagner C., Müller H.W., Knorr K. Faraday waves on a viscoelastic liquid. // Physical Review Letters.– V. 83.– 1999.– P. 308–311.
6. Besson T., Edwards S. Two–frequency parametric excitation of surface waves // Physical Review E.– V. 54.– 1996.– P. 507–513
7. Silber M., Topaz C., Skeldon A. Two–frequency forced Faraday waves: weakly damped modes and pattern selection // Physica D.– V. 143.– 2000.– P. 205–225.
8. Levich V.G. Physicichemical Hydrodynamics.- Prentice-Hall, Engwood Cliff.- 1962.
9. Wang Q., Feder A., Mazur E. Capillary wave damping in heterogenous monolayers // Journal of Physical Chemistry.– V. 98.– № 48.– 1994.– P. 12720–12726.
10. Landau L.D., Lifshits E.M. Fluid Mechsnics.- Oxford, London.- 1987.